\documentclass[11pt]{article}

\usepackage{amssymb}
\usepackage[dvips]{lscape,graphicx}
\include{epsf}
\include{psfig}
\include{epsfig}
\usepackage{epsfig}
\newcommand{\bc}{\begin{center}}
\newcommand{\ec}{\end{center}}
\newcommand{\bd}{\begin{displaymath}}
\newcommand{\ed}{\end{displaymath}}
\newcommand{\be}{\begin{equation}}
\newcommand{\ee}{\end{equation}}
\newcommand{\ba}{\begin{array}}
\newcommand{\ea}{\end{array}}
\newcommand{\bt}{\begin{tabular}}
\newcommand{\et}{\end{tabular}}

\newcommand{\ds}{\displaystyle}

\begin{document}

\thispagestyle{empty}
\begin{flushright}
GUTPA/04/12/01
\end{flushright}
\renewcommand{\thefootnote}{\fnsymbol{footnote}}
\begin{center}{\LARGE{{\bf The Two-Higgs Doublet Model and the 
Multiple Point Principle\footnote{\it To be published in the 
Proceedings of the Bled Workshop on 
What comes beyond the Standard Model? 19 - 30 July 2004.} }}}
\end{center}
\begin{center}{\bf \large{C. D. Froggatt}}
%~\footnote[2]{E-mail: c.froggatt@physics.gla.ac.uk}\\}
\\
\vspace{-4mm}
\end{center}
\renewcommand{\thefootnote}{\arabic{footnote}}
\begin{center}{{\it Department of Physics and Astronomy}\\{\it
University of Glasgow, Glasgow G12 8QQ, Scotland}}\end{center}
\begin{center}{\bf \large{L.V. Laperashvili}}
%~\footnote[1]{E-mail: hbech@alf.nbi.dk}\\}
\\
\vspace{-4mm}
\end{center}
\renewcommand{\thefootnote}{\arabic{footnote}}
\begin{center}{\it  Theory Department, ITEP, Moscow, Russia}
\end{center}
\begin{center}{\bf \large{R. B.
Nevzorov}}
%~\footnote[1]{E-mail: hbech@alf.nbi.dk}\\}
\\
\vspace{-4mm}
\end{center}
\renewcommand{\thefootnote}{\arabic{footnote}}
\begin{center}{{\it School of Physics and Astronomy, 
University of Southampton, UK}\\{\it and Theory Department, ITEP, 
Moscow, Russia}}\end{center}
\begin{center}{\bf \large{H. B.
Nielsen}}
%~\footnote[1]{E-mail: hbech@alf.nbi.dk}\\}
\\
\vspace{-4mm}
\end{center}
\renewcommand{\thefootnote}{\arabic{footnote}}
\begin{center}{{\it  Niels Bohr Institute, }\\{\it
Blegdamsvej 17-21, DK 2100 Copenhagen, Denmark}}\end{center}
\begin{center}{\bf \large{M. Sher}}
%~\footnote[1]{E-mail: hbech@alf.nbi.dk}\\}
\\
\vspace{-4mm}
\end{center}
\renewcommand{\thefootnote}{\arabic{footnote}}
\begin{center}{{\it Physics Department, College of William and Mary,}\\
{\it Williamsburg, Virginia, USA 23187}}\end{center}
\setcounter{footnote}{0}

\begin{abstract}
According to the multiple point principle, Nature
adjusts coupling parameters so that many vacuum states exist and
each has approximately zero vacuum energy density. We apply this
principle to the general two-Higgs doublet extension of the
Standard Model, by requiring the existence of a large set of
degenerate vacua at an energy scale much higher than the presently
realized electroweak scale vacuum. It turns out that two scenarios
are allowed. In the first scenario, a CP conserving Higgs
potential and the absence of flavour changing neutral currents are
obtained without fine-tuning. In the second scenario, the photon
becomes massive in the high scale vacua. We briefly discuss the
resulting phenomenology.
\end{abstract}
%%%%%%%%%%%%%%%%%%%%%%%%%%%%%%%%%%%%%%%%%%%%%%%%%%%%%%%
\thispagestyle{empty}

\newpage

\section{Introduction}
\label{introduction}

The success of the Standard Model (SM) strongly supports the concept of
spontaneous $SU(2)\times U(1)$ symmetry breaking. The mechanism of
electroweak symmetry breaking, in its minimal version, requires the
introduction of a single doublet of scalar complex Higgs fields
and leads to the existence of a neutral massive particle --- the Higgs
boson. Over the past two decades the upper \cite{1} and lower
\cite{1}-\cite{2} theoretical bounds on its mass have been
established. Although the Higgs boson still remains elusive, the
combined analysis of electroweak data indicates that its mass lies
below 251 GeV with 95\% confidence level \cite{4A}.
%\cite{3}--\cite{4}.
Recently the experimental lower limit on the Higgs mass of 115.3
GeV was set by the unsuccessful search at LEPII \cite{4}. The
upgraded Tevatron, LHC and LC have a good chance to discover the
Higgs boson in the near future.

There is, of course, no strong argument for the existence of just
a single Higgs doublet, apart from simplicity. Indeed the
symmetries of many models for physics beyond the SM, such as
supersymmetry or the Peccei-Quinn symmetry \cite{21}, naturally
introduce extra Higgs doublets with unit weak hypercharge. In this
paper, we consider the application of the Multiple Point
Principle to the general two Higgs doublet model, without any
symmetries imposed beyond those of the SM gauge group.

The Multiple Point Principle (MPP) \cite{11} postulates the
co-existence in Nature of many phases, which are allowed by a
given theory. It corresponds to the special (multiple) point on
the phase diagram of the considered theory where these phases
meet. At the multiple point the vacuum energy densities (the
cosmological constants) of the neighbouring phases are degenerate.
Thus, according to MPP, Nature fine-tunes the couplings to their
values at the multiple point. We have not identified the physical
mechanism underlying MPP, but it seems likely \cite{11} that a
mild form of non-locality is required, due to baby universes say
\cite{14}, as in quantum gravity.

When applied to the pure SM, the MPP exhibits a remarkable agreement
\cite{12} with the top quark mass measurements. According to MPP, the
renormalization group improved SM Higgs effective potential
\be
V_{eff}(\phi)=-m^2(\phi)\phi^2+\ds\frac{\lambda(\phi)}{2}\phi^4\,,
\label{1}
\ee
has two rings of minima with the same vacuum energy density \cite{12}.
The radius of the little ring is equal to the electroweak vacuum expectation
value of the Higgs field $|\phi| = v = 246$ GeV. The second vacuum was
assumed to be near the fundamental scale of the theory\footnote{Here
we assume the existence of the hierarchy $v/M_{Pl} \sim 10^{-17}$.
However some of us have speculated \cite{13} that this huge scale ratio
could be derived from MPP, as a consequence of the existence of yet
another SM vacuum at the electroweak scale, formed by the condensation
of a strongly bound S-wave state of 6 top quarks and 6 anti-top quarks.},
identified as the Planck scale $|\phi| \approx M_{Pl}$.
The mass parameter $m$ in the effective potential (\ref{1}) has to be of the
order of the electroweak scale $v$ and is negligible compared to $M_{Pl}$.
The conditions for a second degenerate minimum of $V_{eff}$ at the
Planck scale then become
\be
\beta_{\lambda}(\lambda(M_{Pl}),g_t(M_{Pl}),g_i(M_{Pl}))
= \ds\frac{d\lambda}{d\ln\phi}(M_{Pl}) = \lambda(M_{Pl}) = 0
\label{sm}
\ee
where $g_i(\phi)$ and $g_t(\phi)$ denote the gauge and top quark Yukawa
couplings respectively. Hence, by virtue of MPP, $\lambda(M_{Pl})$ and
$g_t(M_{Pl})$ are determined and one can compute quite precisely the
predicted top quark (pole) and Higgs boson masses using the renormalization
group flow \cite{12}:
\be
M_t=173\pm 5\ \mbox{GeV}\, ,\qquad M_H=135\pm 9\ \mbox{GeV}\, .
\label{3}
\ee

Here we study the MPP predictions for the general two Higgs
doublet extension of the SM \cite{211},\cite{16}. The structure of
the general two Higgs doublet model is outlined in the next
section. The MPP conditions are then formulated in section 3.
Section 4 contains our conclusions.

\section{Two Higgs doublet extension of the SM}
\label{2HDM}

The most general renormalizable $SU(2)\times U(1)$ gauge invariant potential
of the model involving two Higgs doublets is given by
\be
\begin{array}{c}
V_{eff}(H_1, H_2) = m_1^2(\Phi)H_1^{\dagger}H_1 +
m_2^2(\Phi)H_2^{\dagger}H_2
- \biggl[m_3^2(\Phi) H_1^{\dagger}H_2+h.c.\biggr]+\\[3mm]
\ds\frac{\lambda_1(\Phi)}{2}(H_1^{\dagger}H_1)^2 +
\frac{\lambda_2(\Phi)}{2}(H_2^{\dagger}H_2)^2 +
\lambda_3(\Phi)(H_1^{\dagger}H_1)(H_2^{\dagger}H_2) +
\lambda_4(\Phi)|H_1^{\dagger}H_2|^2\\[3mm]
\ds + \biggl[\frac{\lambda_5(\Phi)}{2}(H_1^{\dagger}H_2)^2 +
\lambda_6(\Phi)(H_1^{\dagger}H_1)(H_1^{\dagger}H_2)+
\lambda_7(\Phi)(H_2^{\dagger}H_2)(H_1^{\dagger}H_2)+h.c. \biggr]
\end{array}
\label{4}
\ee
where
$$
H_n=\left(
\ba{c}
\chi^+_n\\[2mm]
(H_n^0+iA_n^0)/\sqrt{2}
\ea
\right) \qquad n=1,2\,.
$$
It is easy to see that the number of couplings in the two Higgs
doublet model (2HDM) compared with the SM grows from two to ten.
Furthermore, four of them $m_3^2$, $\lambda_5$, $\lambda_6$ and
$\lambda_7$ can be complex, inducing CP--violation in the Higgs
sector. In what follows we suppose that the mass parameters $m_i^2$
and Higgs self--couplings $\lambda_i$ of the effective potential
(\ref{4}) only depend on the overall sum of the squared norms of
the Higgs doublets, i.e.
$$
\Phi^2=\Phi_1^2+\Phi_2^2\,,\qquad \Phi_n^2=H_n^{\dagger}H_n =
\frac{1}{2}\biggl[(H_n^0)^2+(A_n^0)^2\biggr]+|\chi_n^+|^2\,.
$$
The running of these couplings is described by the 2HDM renormalization group
equations \cite{171}--\cite{17}, where the renormalization scale is
replaced by $\Phi$.

At the physical minimum of the scalar potential (\ref{4}) the
Higgs fields develop vacuum expectation values
\be
<\Phi_1>=\ds\frac{v_1}{\sqrt{2}}\,,\qquad\qquad
<\Phi_2>=\ds\frac{v_2}{\sqrt{2}}
\label{41}
\ee
breaking the $SU(2)\times U(1)$ gauge symmetry and
generating masses for the bosons and fermions. Here the overall
Higgs norm $<\Phi>=\sqrt{v_1^2+v_2^2}=v=246\,\mbox{GeV}$ is fixed
by the electroweak scale. At the same time the
ratio of the Higgs vacuum expectation values remains arbitrary.
Hence it is convenient to introduce $\tan\beta=v_2/v_1$.

In general the Yukawa couplings of the quarks to the Higgs fields
$H_1$ and $H_2$ generate phenomenologically unwanted flavour
changing neutral currents, unless there is a protecting custodial
symmetry \cite{19}. Such a custodial symmetry requires the
vanishing of the Higgs couplings $\lambda_6$ and $\lambda_7$. It
also requires the down-type quarks to couple to just one Higgs
doublet, $H_1$ say, while the up-type quarks couple either to
the same Higgs doublet $H_1$ (Model I) or to the second Higgs doublet
$H_2$ (Model II) but not both\footnote{Similarly the leptons are
required to only couple to one Higgs doublet, usually chosen to be the
same as the down-type quarks. However there are variations of Models I and
II, in which the leptons couple to $H_2$ rather than to $H_1$.}. If,
in addition, the Higgs coupling $\lambda_5$ vanishes, as in supersymmetric
and Peccei-Quinn models, there is no CP-violation in the Higgs sector.

We emphasize that, in this paper, we do not impose any custodial
symmetry but rather consider the general Higgs potential
(\ref{4}). Instead we require that at some high energy scale
($M_Z<<\Lambda\lesssim M_{Pl}$), which we shall refer to as the
MPP scale $\Lambda$, a large set of degenerate vacua allowed by
the 2HDM is realized. In compliance with the MPP, these vacua and
the physical one must have the same energy density. Thus the MPP
implies that the couplings $\lambda_i(\Lambda)$ should be adjusted
with an accuracy of order $v^2/\Lambda^2$, in order to arrange an
appropriate cancellation among the quartic terms in the effective
potential (\ref{4}).

\section{Implementation of the MPP in the 2HDM}

In this section, we aim to determine a large set of minima of the
2HDM scalar potential with almost vanishing energy density, which
may exist at the MPP high energy scale $\Lambda$ where the mass
terms in the potential can be neglected. The most general vacuum
configuration takes the form: \be <H_1>=\Phi_1\left( \ba{c}
0\\[2mm]
1
\ea
\right)\,,\qquad
<H_2>=\Phi_2\left(
\ba{c}
\sin\theta\\[2mm]
\cos\theta\, e^{i\omega}
\ea
\right)\,,\\[2mm]
\label{6}
\ee
where $\Phi_1^2+\Phi_2^2=\Lambda^2$. Here, the gauge is fixed so
that only the real part of the lower component of $H_1$ gets a vacuum
expectation value.

We now consider the conditions that must be satisfied in order
that minima of $V_{eff}$ should exist for all possible values
of the phase $\omega$. The $\omega$-dependent part of the
potential takes the form:
\be
V_{\omega} =
\frac{\lambda_5(\Phi)}{2} \Phi_1^2 \Phi_2^2 \cos^2\theta
e^{2i\omega} +\left[ \lambda_6(\Phi) \Phi_1^3 \Phi_2
+\lambda_7(\Phi) \Phi_1 \Phi_2^3  \right]\cos\theta e^{i\omega}
+h.c.
\label{Vomega}
\ee
In order that $V_{\omega}$ should become
independent of $\omega$ at the MPP scale minima, we require that
the coefficients of $e^{i\omega}$ and $e^{2i\omega}$ in
(\ref{Vomega}) both vanish at $\Phi = \Lambda$. Similarly for
minima to exist for all values of $\omega$, we require the
derivatives:
\begin{eqnarray}
\lefteqn{\frac{\partial V_{\omega}}{\partial \Phi_1} =
\left[ \lambda_5(\Phi) \Phi_1 \Phi_2^2  +
\beta_{\lambda_5}(\Phi) \frac{\Phi_1^3 \Phi_2^2}{2\Phi^2}
\right] \cos^2\theta e^{2i\omega}  +
}  \nonumber \\
& & \left[3\lambda_6 \Phi_1^2 \Phi_2   +
\beta_{\lambda_6} \frac{\Phi_1^4 \Phi_2}{\Phi^2}  +
\lambda_7 \Phi_2^3  +
\beta_{\lambda_7} \frac{\Phi_1^2 \Phi_2^3}{\Phi^2}
\right]\cos\theta e^{i\omega}  +h.c.
%\nonumber\\& &  + h.c.
\label{dV1}
\end{eqnarray}
and
\begin{eqnarray}
\lefteqn{\frac{\partial V_{\omega}}{\partial \Phi_2} =
\left[ \lambda_5(\Phi) \Phi_1^2 \Phi_2  +
\beta_{\lambda_5}(\Phi) \frac{\Phi_1^2 \Phi_2^3}{2\Phi^2}
\right]\cos^2\theta e^{2i\omega}  + } \nonumber \\
& & \left[\lambda_6 \Phi_1^3   +
\beta_{\lambda_6} \frac{\Phi_1^3 \Phi_2^2}{\Phi^2}  +
3\lambda_7 \Phi_2^2 \Phi_1  +
\beta_{\lambda_7} \frac{\Phi_1 \Phi_2^4}{\Phi^2}
\right]\cos\theta e^{i\omega} +h.c.
%\nonumber \\& &  + h.c.
\label{dV2}
\end{eqnarray}
to be independent of $\omega$ at $\Phi = \Lambda$. Here
$\beta_{\lambda_i}(\Phi) = \frac{d \lambda_i}{d \ln \Phi}(\Phi)$
is the renormalisation group beta function for the Higgs self-coupling
$\lambda_i(\Phi)$.

It is readily verified (unless $\cos\theta$ = 0) that the
vanishing of the coefficients of $e^{i\omega}$ and $e^{2i\omega}$
in Eqs.~(\ref{Vomega}) - (\ref{dV2}) leads to the conditions:
\be
\lambda_5(\Lambda)=\lambda_6(\Lambda) = \lambda_7(\Lambda) = 0
\label{lambda0}
\ee
and
\be
 \beta_{\lambda_5}(\Lambda) = \beta_{\lambda_6}\Phi_1^2 +
 \beta_{\lambda_7}\Phi_2^2 = 0. \label{betalambda0}
 \ee
 When
$\lambda_5 = \lambda_6 = \lambda_7 = 0$, the Higgs potential
manifests an extra Peccei-Quinn-like $U(1)$ symmetry, and the only
non-vanishing contributions to the beta functions
$\beta_{\lambda_5}$, $\beta_{\lambda_6}$ and $\beta_{\lambda_7}$
arise from the Yukawa couplings to the fermion sector. We shall
consider just the third generation fermions here and neglect the
smaller Yukawa couplings from the first two generations. An
obvious method of ensuring that $\beta_{\lambda_5}$,
$\beta_{\lambda_6}$ and $\beta_{\lambda_7}$ also vanish, and
thereby satisfy Eq.~(\ref{betalambda0}), is to extend the $U(1)$
symmetry to the fermion sector at the MPP scale. In other words
the Yukawa couplings at the MPP scale can be taken to be of the
2HDM Model I or Model II form discussed in section \ref{2HDM} This
is illustrated by the explicit expression for $\beta_{\lambda_5}$
(in a notation where we have re-defined the Higgs doublets so that
the top quark only couples to $H_2$ at the MPP scale): \be
 \beta_{\lambda_5}\biggl( \lambda_5=\lambda_6=\lambda_7=0, \Lambda
\biggr)= -\frac{1}{(4\pi)^2}\biggl[12h_b^2(\Lambda)g_b^2(\Lambda)+
4h_{\tau}^2(\Lambda) g^{\ast 2}_{\tau}(\Lambda)\biggr]\,.
\label{34}
\ee
Here $h_b$ and $g_b$ are the couplings of $H_1$ and
$H_2$ to the $b$--quark, while $h_{\tau}$ and $g_{\tau}$ are the
corresponding couplings of the Higgs doublets to the
$\tau$--lepton. For definiteness, we have chosen a phase
convention in which $h_t$, $h_b$, $h_{\tau}$ and $g_b$ are real
and $g_{\tau}$ is complex. The beta function (\ref{34}) vanishes
when
\be
\ba{clcl} (I)\quad&
h_b(\Lambda)=h_{\tau}(\Lambda)=0\,;&\qquad\qquad (II)\quad &
g_b(\Lambda)=g_{\tau}(\Lambda)=0\,;\\
(III)\quad& h_b(\Lambda)=g_{\tau}(\Lambda)=0\,;&\qquad\qquad
(IV)\quad & g_b(\Lambda)=h_{\tau}(\Lambda)=0\,.
\ea \label{35}
\ee
corresponding to the 2HDM Model I and Model II Yukawa couplings
and their leptonic variations.

An alternative method of solving the MPP conditions
(\ref{lambda0}, \ref{betalambda0}), without a
Peccei-Quinn-like $U(1)$ symmetry, is for the $b$ and $\tau$
contributions to cancel in Eq.~(\ref{34})
with $g_{\tau}$ being imaginary. However the
manifold of such MPP solutions in the space of coupling constants
is of the same dimension as that of the Peccei-Quinn-like solutions.
Hence no fine-tuning is required to obtain one of the
Peccei-Quinn-like MPP solutions, as they are just as abundant as
MPP solutions without such a $U(1)$ symmetry. We shall therefore
concentrate on the phenomenologically favoured MPP solutions,
having the 2HDM Model I or Model II Yukawa couplings.

The Peccei-Quinn-like $U(1)$ custodial symmetry of the Higgs and
Yukawa sector implies that
\be
\beta_{\lambda_5}(\Lambda)
= \beta_{\lambda_6}(\Lambda) = \beta_{\lambda_7}(\Lambda) = 0.
\label{beta0}
\ee
It then follows from Eqs.~(\ref{lambda0}) and
(\ref{beta0}) that the renormalization group evolution does not
generate any $U(1)$ custodial symmetry breaking couplings below
the MPP scale, where we thus have:
\be
\lambda_5(\Phi) =
\lambda_6(\Phi) = \lambda_7(\Phi)=0.
\label{lambda567}
\ee
In this
way we have naturally obtained the absence of flavour changing neutral
currents and a CP conserving Higgs potential from the MPP
requirement that vacua at the MPP scale should be degenerate with
respect to the phase $\omega$.

We now consider whether we can impose further MPP conditions on
the couplings. The simplest way to ensure that the quartic part of
the effective potential vanishes for any vacuum configuration
(\ref{6}) at the MPP scale is to impose the condition that all the
self-couplings should vanish there:
\be
\lambda_1(\Lambda)=\lambda_2(\Lambda)=\lambda_3(\Lambda)=
\lambda_4(\Lambda)=\lambda_5(\Lambda)=\lambda_6(\Lambda)=
\lambda_7(\Lambda)=0\,.
\label{5}
\ee
However, further
investigation reveals that the configurations (\ref{6}) do not
correspond to minima of the effective potential in this case. This
can be shown by consideration of the 2HDM renormalization group
equations for the quartic couplings \cite{171}--\cite{17}. The
detailed results depend on the choice of Model I or Model II
Yukawa couplings, but they are qualitatively similar. So, for
convenience, we shall concentrate on the Model II couplings here.

\begin{figure}[t]
\leavevmode
%\vspace{-4cm}
\centerline{ \epsfxsize=12cm \epsfbox{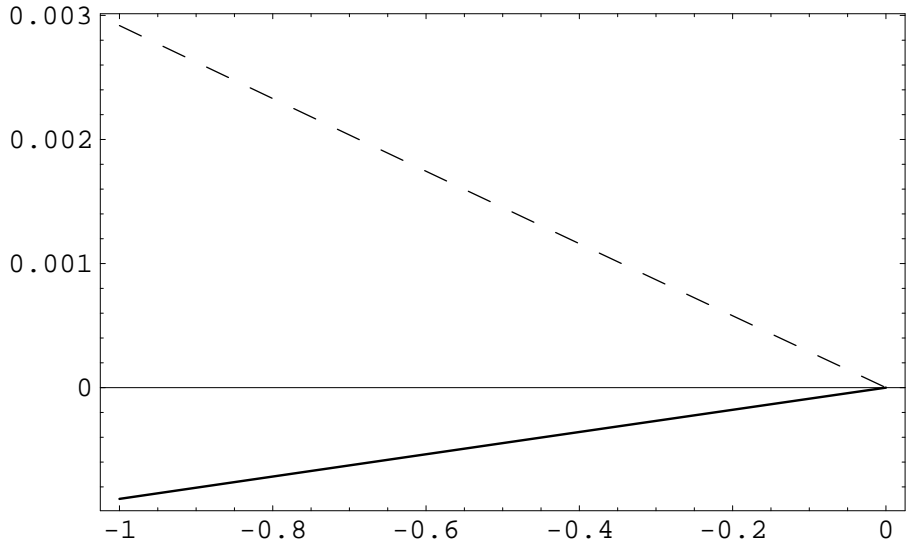} }
\caption{The running of $\lambda_1$ and $\lambda_2$,
as a function of $\log[\Phi^2/M_{Pl}^2]$,
below $M_{Pl}$ for $\lambda_i(M_{Pl})=0$,
$m_t(M_t)=165\,\mbox{GeV}$ and $\alpha_3(M_Z)=0.117$. The
renormalization group flow is plotted for $\tan\beta=2$.
The solid and dashed lines correspond to $\lambda_1$
and $\lambda_2$ respectively.} \label{fig:higgs1}
\end{figure}

\begin{figure}[t]
\leavevmode
%\vspace{-4cm}
\centerline{ \epsfxsize=12cm \epsfbox{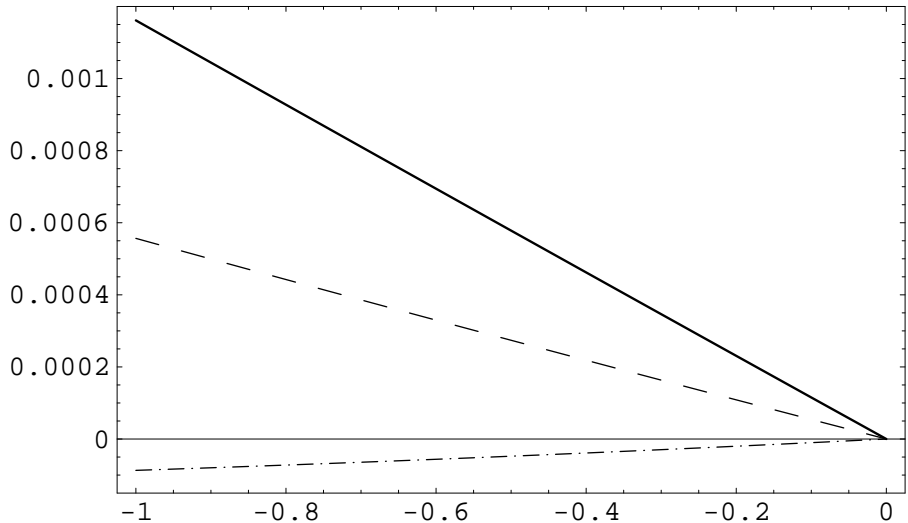} }
\caption{The running of $\lambda_1$, $\lambda_2$ and
$\tilde{\lambda}$, as a function of $\log[\Phi^2/M_{Pl}^2]$,
below $M_{Pl}$ for $\lambda_i(M_{Pl})=0$,
$m_t(M_t)=165\,\mbox{GeV}$ and $\alpha_3(M_Z)=0.117$. The
renormalization group flow is plotted for $\tan\beta=50$.
The solid, dashed and dash--dotted lines correspond to
$\lambda_1$, $\lambda_2$ and $\tilde{\lambda}$
respectively.} \label{fig:higgs2}
\end{figure}

For moderate values of $\tan\beta$ the Higgs self--coupling
$\lambda_1$ becomes negative just below
the MPP scale (see Fig.~\ref{fig:higgs1}).
The renormalization group running of
$\lambda_2$ exhibits the opposite behaviour, because of the large
and negative top quark contribution to the corresponding
beta--function. This means that $V_{eff}$ does not have a
minimum at the
MPP scale and, just below it, there is a  huge negative energy density
$(V_{eff}\sim -\Lambda^4)$ where $ <\Phi_2>=0$ and
$<\Phi_1>\lesssim \Lambda$.

The renormalization group flow of $\lambda_1$
changes at very large $\tan\beta$ (see Fig.~\ref{fig:higgs2}).
The absolute
value of the $b$--quark and $\tau$--lepton contribution to
$\beta_{\lambda_1}$, although negligible at moderate values of
$\tan\beta$, grows with increasing $\tan\beta$. At $\tan\beta\sim
m_t(M_t)/m_b(M_t)$ their negative contribution to the
beta function of $\lambda_1$ prevails over the positive contributions
coming from loops containing Higgs and gauge bosons. The
negative sign of $\beta_{\lambda_1}$ results in
$\lambda_1(\Phi)>0$ if the overall Higgs norm $\Phi$ is less than
$\Lambda$.
However the positive sign of $\lambda_1$ does not ensure the
stability of the vacua (\ref{6}). Substituting the vacuum
configuration (\ref{6}) into the quartic part of the 2HDM scalar
potential, and using Eq.~(\ref{lambda567}), one
finds for any $\Phi$ below the MPP scale:
\be
\ba{rcl}
V(H_1,H_2)&\approx&\ds\frac{1}{2}\biggl(\sqrt{\lambda_1(\Phi)}\Phi_1^2-
\sqrt{\lambda_2(\Phi)}\Phi_2^2\biggr)^2+\\[3mm]
&&+\left(\sqrt{\lambda_1(\Phi)\lambda_2(\Phi)}+\lambda_3(\Phi)+
\lambda_4(\Phi)\cos^2\theta\right) \Phi_1^2\Phi_2^2\, .
\ea
\label{8}
\ee
The Higgs scalar potential (\ref{8})
attains its minimal value for $\cos\theta=0$ if $\lambda_4>0$ or
$\cos\theta=\pm 1$ when $\lambda_4<0$. For these values of $\cos \theta$,
the scalar potential can be written as
\be
V_{eff}(H_1,H_2)\approx\ds\frac{1}{2}\biggl(\sqrt{\lambda_1(\Phi)}\Phi_1^2-
\sqrt{\lambda_2(\Phi)}\Phi_2^2\biggr)^2+
\tilde{\lambda}(\Phi)\Phi_1^2\Phi_2^2\,,
\label{9}
\ee
where
$$
\tilde{\lambda}(\Phi)=\sqrt{\lambda_1(\Phi)\lambda_2(\Phi)}+\lambda_3(\Phi)+
\min\{0,\lambda_4(\Phi)\}\,.
$$
If at some intermediate scale the combination of the Higgs
self--couplings $\tilde{\lambda}(\Phi)$ is
less than zero, then there exists a minimum with negative
energy density that causes the instability of
the vacua at the electroweak and MPP scales. Otherwise
the Higgs effective potential is positive definite
and the considered vacua are stable.

In Fig.~\ref{fig:higgs2} the Higgs
self--couplings $\lambda_1(\Phi)$ and
$\lambda_2(\Phi)$, as well as $\tilde{\lambda}(\Phi)$,
are plotted as a function of $\Phi$ for a very large value of
$\tan\beta$. It is clear
that the vacuum stability conditions, i.e.
\be
\lambda_1(\Phi)\gtrsim 0\,,\qquad\qquad\lambda_2(\Phi)\gtrsim 0\,,
\qquad\qquad\tilde{\lambda}(\Phi)\gtrsim 0
\label{10}
\ee
are not fulfilled simultaneously. The value of $\tilde{\lambda}(\Phi)$
becomes negative for $\Phi<\Lambda$. So we conclude that indeed the
conditions (\ref{5}) can not provide
a self--consistent realization of the MPP in the 2HDM.

At the next stage it is worth relaxing the conditions (\ref{5}),
by permitting $\lambda_1(\Lambda)$, $\lambda_2(\Lambda)$
and $\lambda_3(\Lambda)$ to take on non--zero values. In order to
avoid a huge and negative vacuum energy density in the
global minimum of the 2HDM effective potential that precludes
the implementation of MPP, the vacuum stability conditions
(\ref{10}) should be satisfied for any $\Phi$ in the interval:
$v\lesssim \Phi\lesssim \Lambda$. In this case
both terms in the quartic part of the scalar potential (\ref{8})
are positive. In order to achieve degeneracy of the vacua
at the electroweak and MPP scales, they must go to zero
separately at the scale $\Lambda$. Since $\lambda_4(\Lambda)$ is still
taken to be zero, the second term in Eq.~(\ref{8}) vanishes when
\be
\lambda_3(\Lambda)\simeq-\sqrt{\lambda_1(\Lambda)\lambda_2(\Lambda)}\,.
\label{11}
\ee
For finite values of $\lambda_1(\Lambda)$ and $\lambda_2(\Lambda)$
the first term in the quartic part of
the scalar potential can also be eliminated by the appropriate choice of
Higgs vacuum expectation values:
\be
\Phi_1=\Lambda\cos\gamma\,,\qquad \Phi_2=\Lambda\sin\gamma\,,\qquad
\tan\gamma=\Biggl(\ds\frac{\lambda_1(\Lambda)}{\lambda_2(\Lambda)}
\Biggr)^{1/4}\,.
\label{12}
\ee
The sum of the quartic terms in $V_{eff}(H_1,H_2)$ then tend to
zero at the MPP scale independently of the angle $\theta$ and the
phase $\omega$.

Nevertheless the situation is not as promising as it first appears,
since again we can show it does not correspond to a local minimum of
$V_{eff}$ at $\Phi = \Lambda$, in which all partial derivatives of the 2HDM
scalar potential go to zero. The degeneracy of the vacua, parameterized
by Eqs.~(\ref{6}) and (\ref{12}), implies that the following derivatives
\be
\ds\frac{\partial V_{eff}(H_1,H_2)}{\partial \Phi_i}\propto
\ds\frac{1}{2}\beta_{\lambda_1}\tan^{-2}\gamma+
\frac{1}{2}\beta_{\lambda_2}\tan^2\gamma+\beta_{\lambda_3}+
\beta_{\lambda_4}\cos^2\theta\,.
\label{13}
\ee
should vanish at the MPP scale for any choice of $\theta$
and $\omega$. In order for these derivatives to be independent of
$\theta$, we require $\beta_{\lambda_4}(\Lambda)=0$. However,
for $\lambda_4(\Lambda)=0$, this requirement is in conflict with the
form of the beta function:
\be
\beta_{\lambda_4}(\Lambda) =
\ds\frac{1}{(4\pi)^2}\biggl[3g_2^2(\Lambda)g_1^2(\Lambda)
+12h_t^2(\Lambda)h_b^2(\Lambda)\biggr]
\label{14}
\ee
which is strictly positive. Thus our attempt to
adapt the MPP idea to the 2HDM with $\lambda_4(\Lambda)=0$ fails.
Then we have two MPP  scenarios.

When $\lambda_4(\Lambda) < 0$ (the first scenario), a
self--consistent implementation of the MPP can only be obtained if
$\lambda_1(\Lambda)$, $\lambda_2(\Lambda)$ and
$\lambda_3(\Lambda)$ have non--zero values. Then $\cos \theta =
\pm1$ near the MPP scale minima, where the Higgs effective
potential takes the form (\ref{9}). In order to ensure the
vanishing of $V_{eff}$ at the MPP scale with an accuracy of order
$v^2/\Lambda^2$, the combination of Higgs self--couplings
$\tilde{\lambda}(\Lambda)$ must go to zero. Furthermore, if the
2HDM effective potential is to possess a set of local minima at
the MPP scale, the derivative of $\tilde{\lambda}(\Phi)$ must
vanish when $\Phi = \Lambda$ and hence
$\beta_{\tilde{\lambda}}(\Lambda) = 0$. In this first scenario,
the following set of MPP scale vacua \be <H_1>=\left(
\begin{array}{c}
0\\ \Phi_1
\end{array}
\right)\; , \qquad <H_2>=\left(
\begin{array}{c}
0\\ \Phi_2\, e^{i\omega}
\end{array}
\right) \label{16} \ee have the same energy density for any
$\omega$. The ratio of the Higgs field norms $\Phi_1$ and $\Phi_2$
in Eq.~(\ref{16}) is defined by the equations for the extrema of
the 2HDM scalar potential, whose solution is given by
Eq.~(\ref{12}). In the minima (\ref{16}) the photon remains
massless and electric charge is conserved.

In the second scenario $\lambda_4(\Lambda)>0$ the parameter
$\cos\theta$ tends to zero. We note that our general derivation of
the MPP conditions (\ref{lambda0}, \ref{betalambda0}), and the
consequent $U(1)$ custodial symmetry without fine-tuning, breaks
down in this case, since the Higgs potential does not depend on
the phase $\omega$ near its minimum (where $\cos \theta =0$). If
$\lambda_4(\Lambda)-|\lambda_5(\Lambda)|>0$ and
$$
\lambda_1(\Lambda)=\lambda_2(\Lambda)=\lambda_3(\Lambda)=
\lambda_6(\Lambda)=\lambda_7(\Lambda)=0
$$
the following set of vacua
\be
<H_1>=\left(
\begin{array}{c}
0\\ \Phi_1
\end{array}
\right)\; , \qquad <H_2>=\left(
\begin{array}{c}
\Phi_2\\ 0
\end{array}
\right) \label{15} \ee are degenerate for any $\Phi_1$ and
$\Phi_2$ satisfying $\Phi_1^2+\Phi_2^2=\Lambda^2$. In order to
ensure the existence of the minima given by Eq.~(\ref{15}), the
conditions for extrema must be fulfilled which lead to
$\beta_{\lambda_1}(\Lambda)=\beta_{\lambda_2}(\Lambda)
=\beta_{\lambda_3}(\Lambda)=0$. The cancellation of different
contributions to these $\beta$-functions only becomes possible for
large values of the Yukawa couplings at the MPP scale
(corresponding to large $\tan\beta$ in Model II). The resulting
vacuum energy density vanishes because $\cos\theta$ goes to zero
in these vacua. At the set of minima (\ref{15}) the $SU(2)\times
U(1)$ gauge symmetry is broken completely and the photon gains a
mass of the order of $\Lambda$. This is not in conflict with
phenomenology, since an MPP scale minimum is not presently
realised in Nature. However, on phenomenological grounds, we
prefer the first scenario, although it is consistent to impose an
{\it ad hoc} $Z_2$ custodial symmetry on the second scenario which
we will discuss separately elsewhere.

The conditions
\be
\left\{
\ba{l}
\lambda_5(\Lambda)=\lambda_6(\Lambda)=\lambda_7(\Lambda)
=\beta_{\lambda_5}(\Lambda) =\beta_{\lambda_6}(\Lambda)
=\beta_{\lambda_7}(\Lambda) = 0 \,,\\[2mm]
\tilde{\lambda}(\Lambda)= \beta_{\tilde{\lambda}}(\Lambda) = 0\,,
%\ds\frac{d\tilde{\lambda}(\Phi)}{d\Phi}\Biggl|_{\Phi=\Lambda}=0
\ea
\right.
\label{17}
\ee
leading to the appearance of the degenerate vacua (\ref{16}) in our
preferred scenario, should be identified
with the MPP conditions analogous to those of Eq.~(\ref{sm}).
The conditions (\ref{17}) have to be supplemented
by the vacuum stability requirements (\ref{10}), which must be valid
everywhere from the electroweak to the MPP scale. Any failure of either
the conditions (\ref{17}) or the inequalities (\ref{10}) prevents a
consistent realization of the MPP in the 2HDM.

We have made a detailed numerical analysis of these MPP constraints on the
Higgs spectrum in the 2HDM for high energy scales ranging from
$\Lambda = M_{Pl}$ down to $\Lambda = 10$ TeV, which we shall report
elsewhere. In the large $\tan\beta$ limit, the allowed range of the
Higgs self--couplings is severely constrained by the MPP conditions
(\ref{17}) and vacuum stability requirements (\ref{10}).
As a consequence, using the Model II Yukawa couplings and the
lower limit on the charged scalar mass \cite{24} deduced from the
non-observation of $B\to X_s\gamma$ decay, the Higgs spectrum exhibits a
hierarchical structure for most of the large $\tan\beta$
($\tan\beta\gtrsim 2$) region. While the heavy scalar, pseudoscalar and
charged Higgs particles are nearly degenerate with a mass greater than
$300\,\mbox{GeV}$, the mass of the SM--like Higgs boson $m_h$ does not
exceed $180\,\mbox{GeV}$ for any scale $\Lambda\gtrsim 10\,\mbox{TeV}$.
The bounds on $m_h$ become stronger as the MPP scale is increased. For
$\Lambda=M_{Pl}$ and large values of $\tan\beta$ we find:
$m_h=137\pm 12\,\,\mbox{GeV}$. However, for very large
$\tan\beta\simeq m_t(M_t)/m_b(M_t)$ the MPP restrictions on the Higgs
self--couplings and the lightest Higgs scalar mass turn out to be
substantially relaxed, due to a loosening of the allowed upper limit
on $\lambda_2(\Lambda)$. In particular the upper bound on $m_h$ is
increased to 180 GeV for $\Lambda = M_{Pl}$ and very large $\tan\beta$.

\section{Conclusion}

We have studied the constraints imposed by the Multiple Point
Principle on the general two Higgs doublet model, by requiring the
existence of a large number of vacua at a high energy scale
$\Lambda$ which are degenerate with the electroweak scale vacuum.
The MPP conditions at the scale $\Lambda$, derived in our
preferred scenario with the vacua (\ref{16}), are summarized in
Eq.~(\ref{17}). In addition the vacuum stability conditions
(\ref{10}) must be satisfied. The MPP conditions in the first line
of Eq.~(\ref{17}) give CP invariance of the Higgs potential
and the presence of a softly broken (by the
$m^2_3H_1^{\dagger}H_2$ term in $V_{eff}$) $Z_2$ symmetry of the
usual type responsible for the absence of flavour changing neutral
currents without fine-tuning. The $Z_2$ invariance of the 2HDM
Lagrangian is not
spoiled by the renormalization group flow. This means that the MPP
provides an alternative mechanism for the suppression of flavour
changing neutral currents in the 2HDM.

In addition the MPP conditions in the second line of Eq.~(\ref{17})
provide two relationships between the non-zero Higgs self-couplings,
at the scale $\Lambda$, which can in principle be checked when the
masses and couplings of the Higgs bosons are measured at future
colliders. It is interesting to remark that these relationships
are satisfied identically in the minimal supersymmetric standard model
at all high energy scales, where the soft SUSY breaking terms can be
neglected.

In conclusion we have constructed a new simple MPP inspired
non-supersymmetric two Higgs doublet extension of the SM.

\vspace{10mm}
\noindent
{\Large \bf Acknowledgements}
\vspace{2mm}

\noindent
The authors are grateful to E.Boos, I.Ginzburg, S.King,
M.Krawczyk, L.Okun and S.Moretti for fruitful discussions.
RN and HN are indebted to
the DESY Theory Group for hospitality extended to them in
2002 when this project was started. RN is also grateful to
Alfred Toepfer Stiftung for the scholarship and
for the favour disposition during his stay in Hamburg (2001--2002).

%\vspace{2mm}
\noindent
LL and RN would like to acknowledge the support of the
Russian Foundation for Basic Research (RFBR),
projects 00-15-96786, 00-15-96562 and 02-02-17379.
RN was also partly supported by a Grant of the President
of Russia for young scientists (MK--3702.2004.2).
The work of CF was
supported by PPARC grant PPA/G/0/2002/00463 and that of MS by the
National Science Foundation grant PHY0243400.

%\newpage

\end{document}